\documentclass[5p,number,sort&compress]{elsarticle}
\usepackage{lineno,float,graphicx,amssymb,amsmath,times,xspace}
\usepackage{hyperref}
\journal{Nuclear Instruments and Methods A}

\def\cref#1{Chapt.\,\ref{#1}\xspace}
\def\Cref#1{Chapter~\ref{#1}\xspace}
\def\sref#1{Sect.\,\ref{#1}\xspace}

\def\fref#1{Fig.\,\ref{#1}\xspace}

\def\Fref#1{Figure~\ref{#1}\xspace}
\def\tref#1{Table~\ref{#1}\xspace}
\def\eref#1{(\ref{#1})\xspace}

\def\lleft{\textit{left}}
\def\rright{\textit{right}}
\def\LLeft{\textit{Left}}
\def\RRight{\textit{Right}}
\def\ttop{\textit{top}}
\def\bbottom{\textit{bottom}}
\def\TTop{\textit{Top}}
\def\BBottom{\textit{Bottom}}

\def\deg{^\circ}

\begin{document}

\begin{frontmatter}

\title{LORA: A scintillator array for LOFAR to measure extensive air showers}

\author[a]{S. Thoudam\corref{cor1}}
\ead{s.thoudam@astro.ru.nl}
\author[a,b]{S. Buitink}
\author[a]{A. Corstanje}
\author[a]{J.E. Enriquez}
\author[a,c,d]{H. Falcke}
\author[c]{W. Frieswijk}
\author[a,d]{J.R. H\"orandel}
\author[a,e]{A. Horneffer}
\author[a,f]{M. Krause}
\author[a,d]{A. Nelles}
\author[a]{P. Schellart}
\author[b]{O. Scholten}
\author[a]{S. ter Veen}
\author[a]{M. van den Akker}
\cortext[cor1]{Corresponding author}
\address[a]{Department of Astrophysics, IMAPP, Radboud University Nijmegen, P.O. Box 9010, 6500 GL Nijmegen, The Netherlands}
\address[b]{KVI, University of Groningen, 9747 AA Groningen, The Netherlands}
\address[c]{ASTRON, 7990 AA Dwingeloo, The Netherlands}
\address[d]{Nikhef, Science Park Amsterdam, 1098 XG Amsterdam, The Netherlands}
\address[e]{now at: Max Planck Institut f\"ur Radioastronomie, 	Auf dem H\"ugel
            69, 53121 Bonn, Germany}
\address[f]{now at: DESY, Platanenallee 6, 15738 Zeuthen, Germany}

\begin{abstract}
The measurement of the radio emission from extensive air showers, induced by
high-energy cosmic rays is one of the key science projects of the LOFAR radio
telescope.  The LOfar Radboud air shower Array (LORA) has been installed in the
core of LOFAR in the Netherlands.  The main purpose of LORA is to  measure the
properties of air showers and to trigger the read-out of the LOFAR radio
antennas to register extensive air showers. The experimental set-up of the
array of scintillation detectors and its performance are described. 
\end{abstract}

\end{frontmatter}

\section{Introduction}\label{sect-1}
The search for the origin of the highest energy particles in the Universe is a
big challenge in astroparticle physics \cite{nagano-watson, Blumer2009,
Hoerandel2008}. From the experimental point of view, a precise measurement of
the elemental composition of cosmic rays at the highest energies is crucial.
The present work is part of an endeavor to establish a new method to measure
air showers at high energies and determine the mass composition of cosmic rays
with nearly $100\%$ duty cycle: the radio detection of air showers \cite{Jelley1965}. To
contribute to the measurement of radio signals from air showers with the LOFAR
telescope \cite{lofar}, we have installed an air shower array in the LOFAR core.

High-energy cosmic rays impinging onto the atmosphere of the Earth, induce
cascades of secondary particles. The bulk of the charged particles are
electrons and positrons. They are deflected in the magnetic field of the Earth,
while in addition, there is an excess of negative charge.  This yields the
emission of coherent radiation with frequencies of tens of MHz, e.\ g.
\cite{Falcke2003, Scholten2008, Huege2012, ZHAireS}.

The feasibility of quantitative radio measurements of air showers has been
demonstrated with the LOPES experiment (LOFAR prototype station)
\cite{Falcke2005, Hoerandel2009a, Schroder2012}. It has been shown that radio
emission can be detected using low-noise amplifiers and fast digitizers in
combination with sufficient computing power to analyze the registered signals.

Radio emission from air showers is detected with the LOFAR radio telescope in
the framework of the LOFAR key science project \emph{Cosmic Rays}
\cite{schellartnelles}. The LOw Frequency ARray (LOFAR) is a 
digital observatory \cite{lofar}. The main focus of the astronomy
community is to observe the radio Universe in the frequency range of
$(10-240)$~MHz.

More than 40 stations with fields of relatively simple antennas work together
as a digital radio interferometer, i.e.\ the measured signals are digitized
with fast ADCs and correlations are formed in a central processing unit. The
antenna fields are distributed over several countries in Europe with a dense
core in the Netherlands. The latter consists of 24 stations on an area
measuring roughly 5~km$^2$. Each station comprises 96 low-band antennas, simple
inverted V-shaped dipoles, operating in the frequency range of $(10-80)$ MHz.
Each antenna has two dipoles, oriented perpendicular to each other.  In
addition, fields of high band antennas\footnote{The fields comprise 48 antennas
in the Dutch stations and 96 in the European ones.} cover the frequency range
of $(110-240)$ MHz. The signals from the antennas are digitized and stored in a
ring buffer (transient buffer board, TBB). A triggered read-out of these
buffers will send the raw data to a central processing facility.

An ultimate goal is to independently detect radio emission from air showers
with LOFAR. This requires a sophisticated trigger algorithm that analyses the
digitized antenna signals in real time. To assist with the development of the
trigger algorithm and to measure basic air shower parameters, an array of
particle detectors has been at LOFAR.

\begin{figure}[t]
\includegraphics[width=\columnwidth]{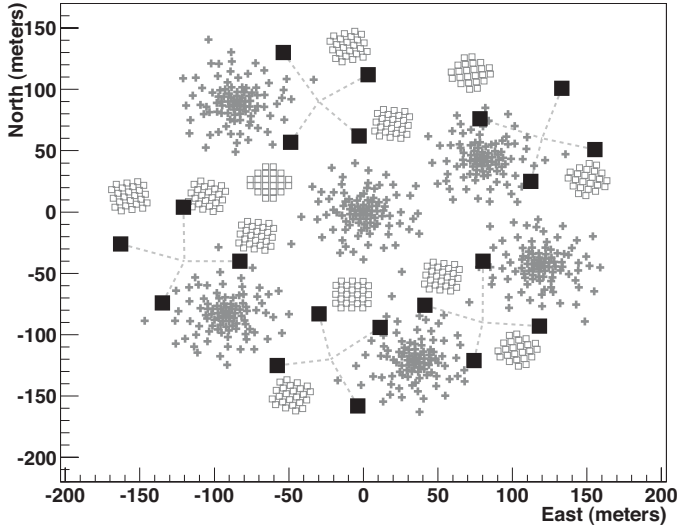}
\caption{\label {lora-layout} Layout of LORA in the dense core in the center of
LOFAR. The squares represent the positions of the particle detectors. The
crosses and open squares represent the two different types of LOFAR radio
antennas.  The dotted lines indicate the grouping of the detectors for the data
acquisition.}
\end{figure}

The LOFAR Radboud Air Shower Array (LORA) is an array of scintillation
counters, located in the innermost center of LOFAR, the \emph{superterp}. It has
been designed to register air showers initiated by primary particles with
energies exceeding $10^{16}$ eV. Strong radio signals are expected from air
showers in this energy region. This energy regime is also of astrophysical
interest, as a transition is expected from a Galactic to an extra-galactic
origin of cosmic rays at energies between $10^{17}$ and $10^{18}$ eV
\cite{Blumer2009, Hoerandel2008}. 

In the following, we describe the set-up of LORA and its properties.  The
experimental set-up is described in \sref{sect-2} and the detector 
calibration in \sref{sect-3}.  The various steps involved in the reconstruction
of air shower parameters are described in \sref{sect-4} and in \sref{sect-5}
the reconstruction accuracies are discussed, followed by a review of the array
performance in \sref{sect-6}.

\section{Experimental set-up}\label{sect-2} 

\begin{figure}[t]
  \includegraphics[width=\columnwidth]{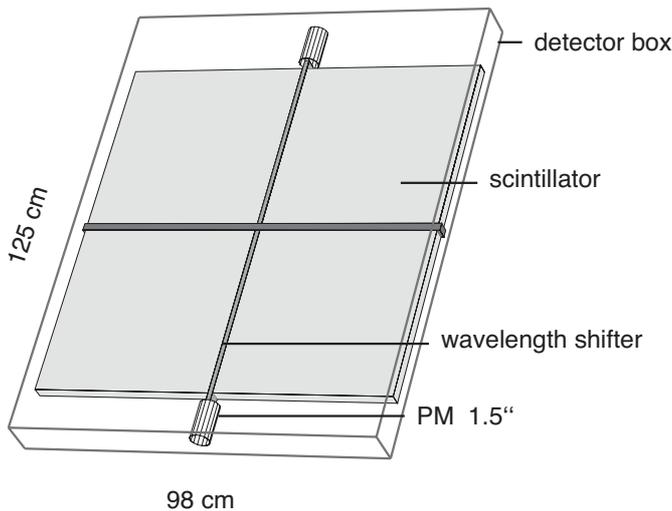}
  \caption{\label{scintillator} Schematic view of a scintillation detector.
       Sheets of plastic scintillator are read out by photomultiplier tubes via
       wavelength shifter bars \cite{kascadenim}.}
\end{figure}

\begin{figure}[t]
  {\centering \includegraphics[width=0.8\columnwidth]{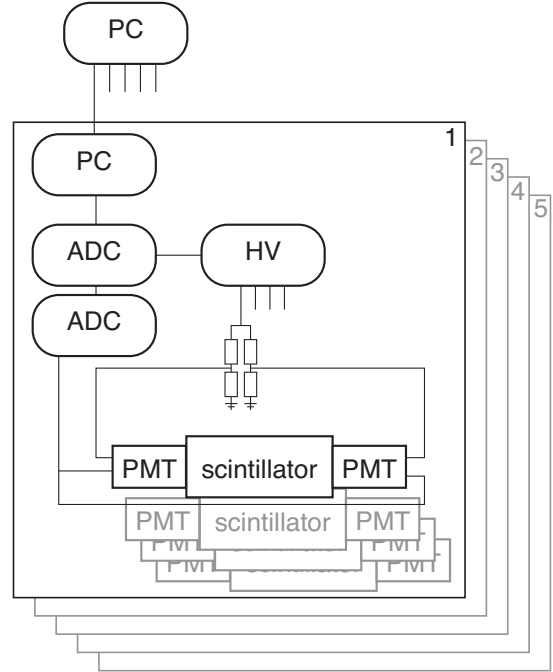}\\}
  \caption{\label{electronics} Schematic view of the electronics 
       components for the data acquisition and experiment control.
       A basic electronics unit serves to read out four scintillator units.
       Five such electronics units are used to read out the twenty detectors.}
\end{figure}

LORA comprises $20$ detector units, located on a circular area with a diameter
of about 320~m.  The positions of the detectors in the innermost core of LOFAR are
illustrated in \fref{lora-layout}.  The array is sub-divided into five units,
each comprising four detectors. The detectors are located on circles with a
radius of about $40$ m around a central electronics unit, with a spacing of
$50$ to $100$~m between the detectors, respectively.

Each detector unit contains two pairs of scintillators (NE $114$) with the
dimensions $47.5 \times 47.5 \times 3$~cm$^3$, read out via wavelength shifter
bars (NE $174$ A) through a photomultiplier tube (EMI $9902$).\footnote{The
detectors were previously operated in the KASCADE calorimeter
\cite{kascadenim}.} A detector unit, containing the two pairs of scintillators
and two photomultiplier tubes  is sketched in \fref{scintillator}.  The
detectors are installed inside weatherproof shelters.

The two photomultipliers in one detector unit share a common high-voltage
channel. To match the gain of the two tubes, we use a resistor network to
adjust the voltage correspondingly. The signals of the two photomultiplier
tubes in each detector are read out via RG223 coaxial cables and a passive connection into a single
digitizer channel. 12-bit ADCs are used, which sample the incoming voltage with
a time resolution of 2.5~ns \footnote{Internally, two ADCs are used per
channel, sampling the same input signal at 200~MHz with an offset of half a
clock cycle.} 
\cite{Hi}. A field programmable gate array (FPGA) provides a trigger signal in
real time. 

Four detectors form (electronically) a unit, comprising two digitizer units
(with two electronic channels each) \cite{Hi, Hi2}, controlled by a Linux-operated,
single-board mini PC.The two digitizer units operate in a master and slave 
combination like a four-channel oscilloscope, where the master generates a 
common trigger for both the digitizer units. 
The master digitizer contains a GPS receiver (Trimble, Resolution T), 
which provides GPS time stamps to both the digitizer units. Each digitizer contains 
a 200 MHz clock counter to assign a time stamp with nanoseconds accuracy to each triggered signal. 

The pulse per second signals (1PPS) from this type of GPS receiver can introduce a timing uncertainty of up to a 
maximum of 20 ns. This error is stored every second during the data taking.
It is corrected for the event time stamp in the offline data analysis using a proper correction formula \cite{HansVK}.
The time stamp calculation also takes into account the fluctuations in the number of clock periods of the 
200 MHz clock counter between two PPS signals. 

The two digitizers are connected to the PC through an USB interface. The PC
also controls a four-channel high-voltage supply through one of the digitizer units.
The FPGA inside the digitizer unit controls an input-output register, which is
connected to the high-voltage supply, allowing to set the individual voltages
on the four channels remotely.  A block diagram of the electronics components
is depicted in \fref{electronics}.

When the four input signals from the PMTs in an electronics unit satisfy a
local trigger condition, usually three out of four detectors in coincidence within 400~ns the digitizers send the data to the local computer.
The data from the five mini PCs consequently are sent via Ethernet to a
central, Linux-operated master computer, where the main data acquisition (DAQ)
runs.  Within 100~ms all data are collected from the other electronics units.
The received time stamps, which are each assigned according to the first threshold crossing in a detector, 
are checked for coincidences (500~ns window) and are combined to an event file that is stored locally.
A simple analysis is performed on these data, which reconstructs
arrival direction and core position to allow for monitoring. In this computer also an additional 
high-level trigger can be formed, based on the number of sub-arrays that have
detected an air shower. This high-level trigger is used to trigger the read-out
of the radio antennas. The overall processing takes about 130 ms (including the
wait time), which is fast compared to the 5~s of data that are stored in the
ring buffers of the radio antennas.

The main DAQ program also controls the DAQ programs running on the mini PCs.
All input parameters, including those required by the DAQ on the local
computers are set on the master computer. The whole DAQ is controlled and
monitored using an online monitoring panel, which can be accessed remotely. The
display panel provides continuous monitoring of the performance of the
electronics and the detectors during operations. Both, the monitoring panel and
the DAQ software use several features from the ROOT package \cite{ROOT}.

\begin{figure}[t]
\includegraphics[width=\columnwidth]{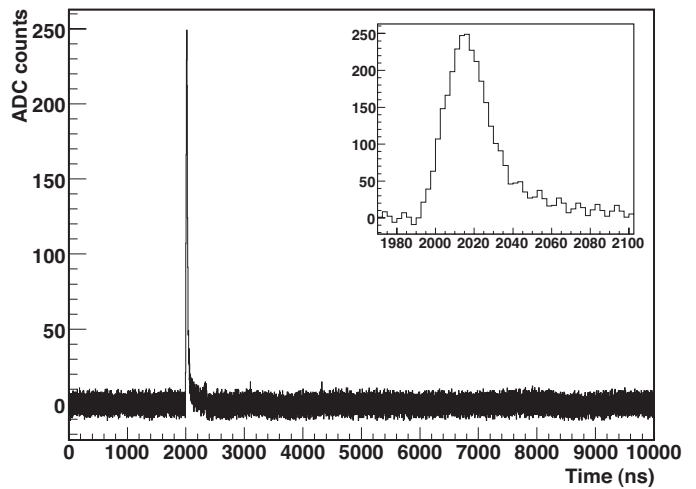}
\caption{\label {traces} Example of a signal time trace, produced by a charged
    particle passing though a detector. The inset shows a closer view
    of the signal around the maximum value (between $1970-2100$ ns).}
\end{figure}

\begin{figure}[t]
\includegraphics[width=\columnwidth]{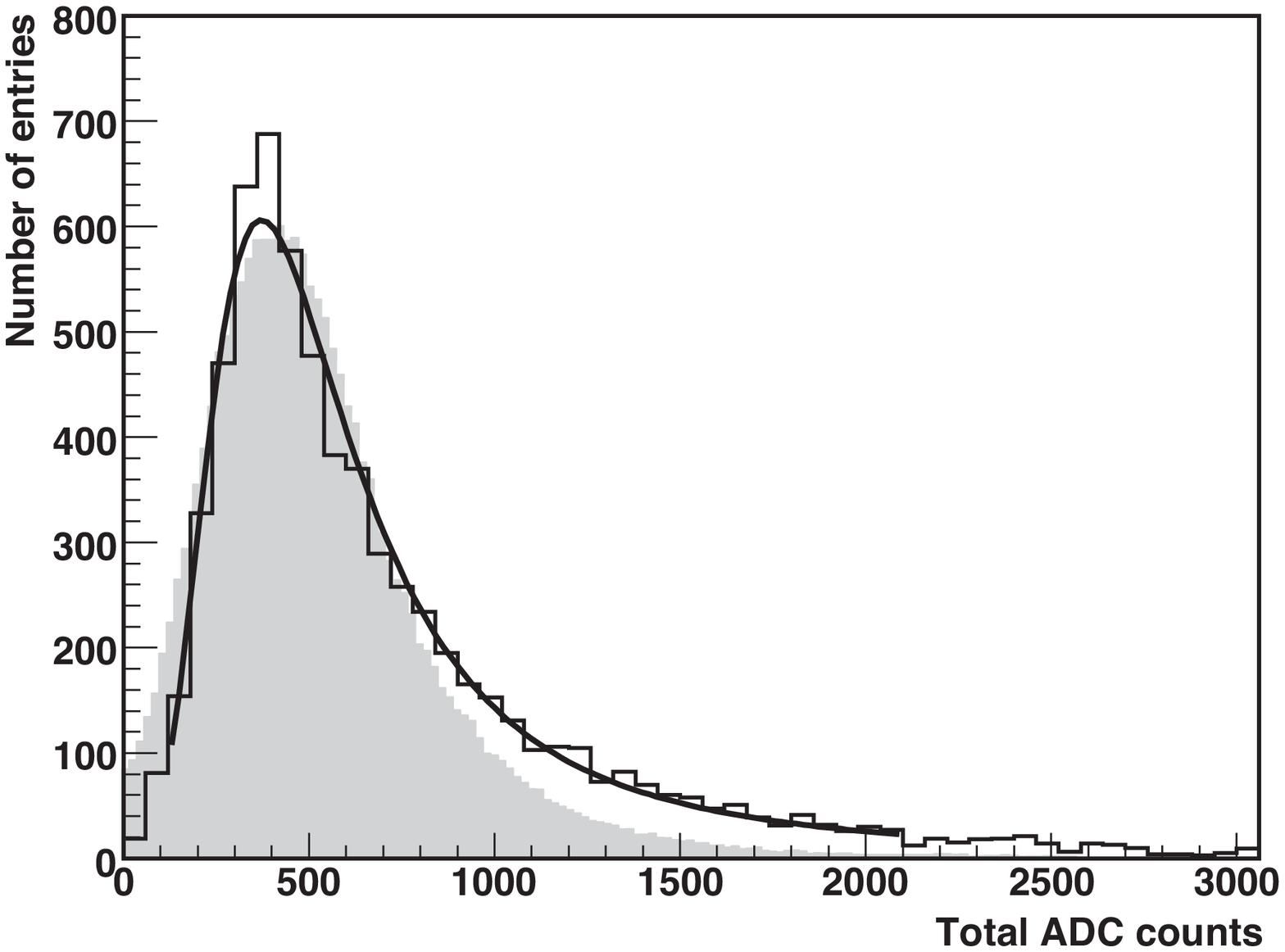}
\includegraphics[width=\columnwidth]{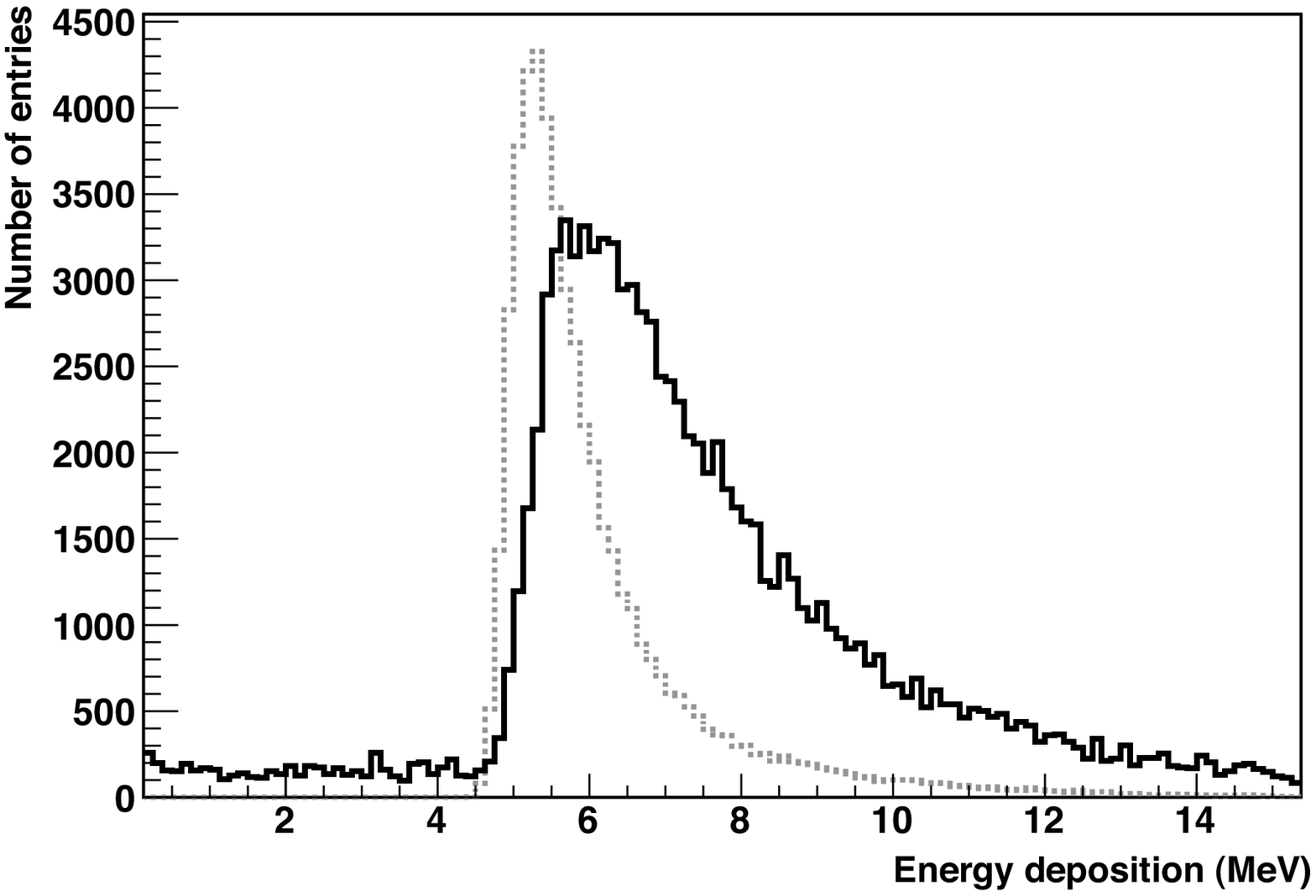}
\caption{\label{landau} Distribution of the total energy deposition by singly
charged particles in a detector. 
\TTop: Measured values, the curve represents a fit of a Landau distribution
function. The most probable value, given by the
fit is $(400.5\pm 3.5)$ ADC counts.
The predictions of a detector simulation, using GEANT4 are indicated
by the shaded histogram.
\BBottom: Results of a simulation, using the GEANT4 code for vertical muons
(dotted histogram) and muons, impinging according to a $\cos(\theta)^2$ zenith
angle distribution (solid histogram).  The most probable energy deposition
amounts to 6.67~MeV.}
\end{figure}

\section{Detector calibration}\label{sect-3}
For each event, traces of the PMT signals are stored in a time window of
$10~\mu$s. We have chosen to start the recorded data $2~\mu$s before the
trigger, thus, we measure ADC traces from $2~\mu$s before to $8~\mu$s after the
trigger for each event. A typical ADC trace is depicted in \fref{traces}.  The
inset shows a closer view around the pulse of a through-going particle.

To calculate the total signal produced by a charged particle (which corresponds
to the total energy deposited by the particle) from the recorded time traces,
the following procedure is applied: The average pedestal is calculated from the
$2~\mu$s window before the trigger. This pedestal is subtracted from the ADC
values and the signal trace is integrated over the time window from
$(t_{peak}-40~\mathrm{ns})$ to $(t_{peak}+250~\mathrm{ns})$. $t_{peak}$ is the
time of the maximum ADC count in the trace.

The resulting measured energy deposition of singly charged particles in a
detector is shown in \fref{landau} (\ttop). A Landau  function is fitted to
the measured distribution. The most probable value corresponds to the energy
deposition of the through-going charged particle.  This value is taken for the
energy calibration of each detector. The high voltage applied to each
photomultiplier is adjusted such that the Landau distribution peaks at $\approx
400$~ADC counts.

In order to determine the energy deposition of singly charged particles, we
conducted simulations with the GEANT4 package \cite{Geant4}.  The scintillators are made of
polyvinyl-toluene (CH$_2$CH(C$_6$H$_4$CH$_3$)$_n$), which is simulated as a C:H
mixture of 9:10, with a density of 1.032 g/cm$^3$. The plates are enclosed in a
lighttight box, made of aluminum plates of 1~mm thickness. Thus, the detectors
are also sensitive to the photon component of air showers, since a significant
fraction of high-energy photons undergoes pair production and thus,
contributes to the signal. The rest of the volume is filled with air.

Muons of 4~GeV are shot through the detector to find the distribution of the
energy deposition. It has been found experimentally that muons arrive under
directions, distributed according to a $\cos^2\theta$ zenith angle
distribution \cite{griederbook}. To get a realistic distribution we take a
random ground location in a square of $1.5\times1.5$~m with a LORA detector
in the center.  A random azimuth angle is chosen, and a zenith angle is chosen
from a $\cos^3\theta \sin \theta$ distribution, where the additional
$\cos\theta\sin\theta$ term is to correct for the acceptance of a flat
detector and the $\sin\theta$-dependency of the solid angle. The resulting deposited energy is shown in \fref{landau} (\bbottom)
for vertically incident muons as well as for muons arriving under a realistic
zenith angle distribution.  As expected, the latter is slightly broader, since
more inclined particles have trajectories larger than the thickness of a
detector (3~cm) and thus, deposit more energy. It also has a plateau at low
energies, caused by particles that hit the edges of the plates and have
trajectories $< 3$~cm.  
The most probable energy deposition amounts to 6.67~MeV. The measured
distribution can be approximated if we take into account the statistical noise,
due to the low number of scintillation photons that induce a signal in the PMT,
and the electronic noise. The resulting distribution is indicated in \fref{landau} (\ttop) as light
grey histogram. Remaining discrepancies can be explained by the missing consideration of the location dependent light deposit in the wave-shifter guide. This will influence the tail of the distribution. It might also affect the position of the peak, however not more than 10\%. 


During operation, the trigger threshold for each individual channel is set with
respect to the corresponding ADC noise of the channel. The recorded noise level
exhibits a dependence on the ambient temperature and in particular, shows
day-night variations. Therefore, we apply a dynamic trigger threshold to the
recorded data. Every hour the threshold is calculated from the noise level,
registered during the last hour. The threshold is set to a value of
$(\bar{N}+4\sigma)$ where $\bar{N}$ and $\sigma$ denote the mean value and the
fluctuation of the noise for the last hour, respectively.

\begin{figure}[t]
 \includegraphics[width=\columnwidth]{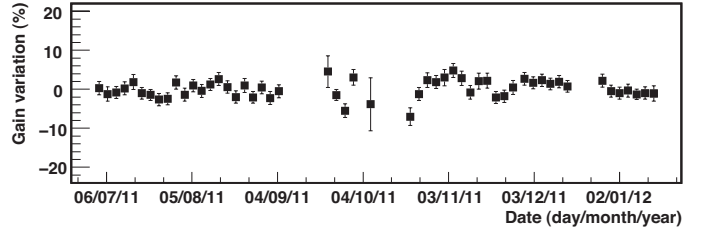}
 \caption{\label{det-gain} Gain of the photomultiplier tubes of a typical
detector as a function of time.} 
\end{figure}

The photomultipliers, placed in weather-proof shelters, are exposed to the
ambient temperature. It is well known that the gain of a photomultiplier
changes as a function of its operating temperature. A stable gain is necessary
for a good performance of the experiment.  The measured gain of a detector
during about 6 months of operation is depicted in \fref{det-gain}.  Each point
represents a value, averaged over three days.  The gain variation is calculated
with respect to the averaged gain. The variation is found to be within
$\pm10\%$.  The overall stability looks good for  18 out of the 20 detectors,
showing gain variations within $\pm 10\%$ for more than $\approx 93\%$ of the
total operation time. The remaining two detectors showed variations up to
$\pm10\%$ within about 70\% and 85\% of the total operation time, respectively.

\section{Data taking}
For data taking, a coincidence trigger condition of 3 out of 4 detectors has
been set for each sub-array. An event is accepted by the master computer if at
least one sub-array has been triggered. These trigger settings generate a total
event rate of $\approx 0.15$~Hz from the full array (five sub-arrays). The
total daily data output amounts to about 180~MB. 

Full operation of LORA started in June 2011. Since then, air showers are
continuously recorded with the set-up. For this article around 162 days of
clean data have been collected with the array. This amounts to a total of  2.1
million air showers.  For the analysis presented below, showers which trigger a
minimum of 5 detectors (at least 2 sub-arrays) are considered. In total, we
have recorded 114659 such air showers.

\begin{figure*}[t]
\includegraphics[width=\columnwidth]{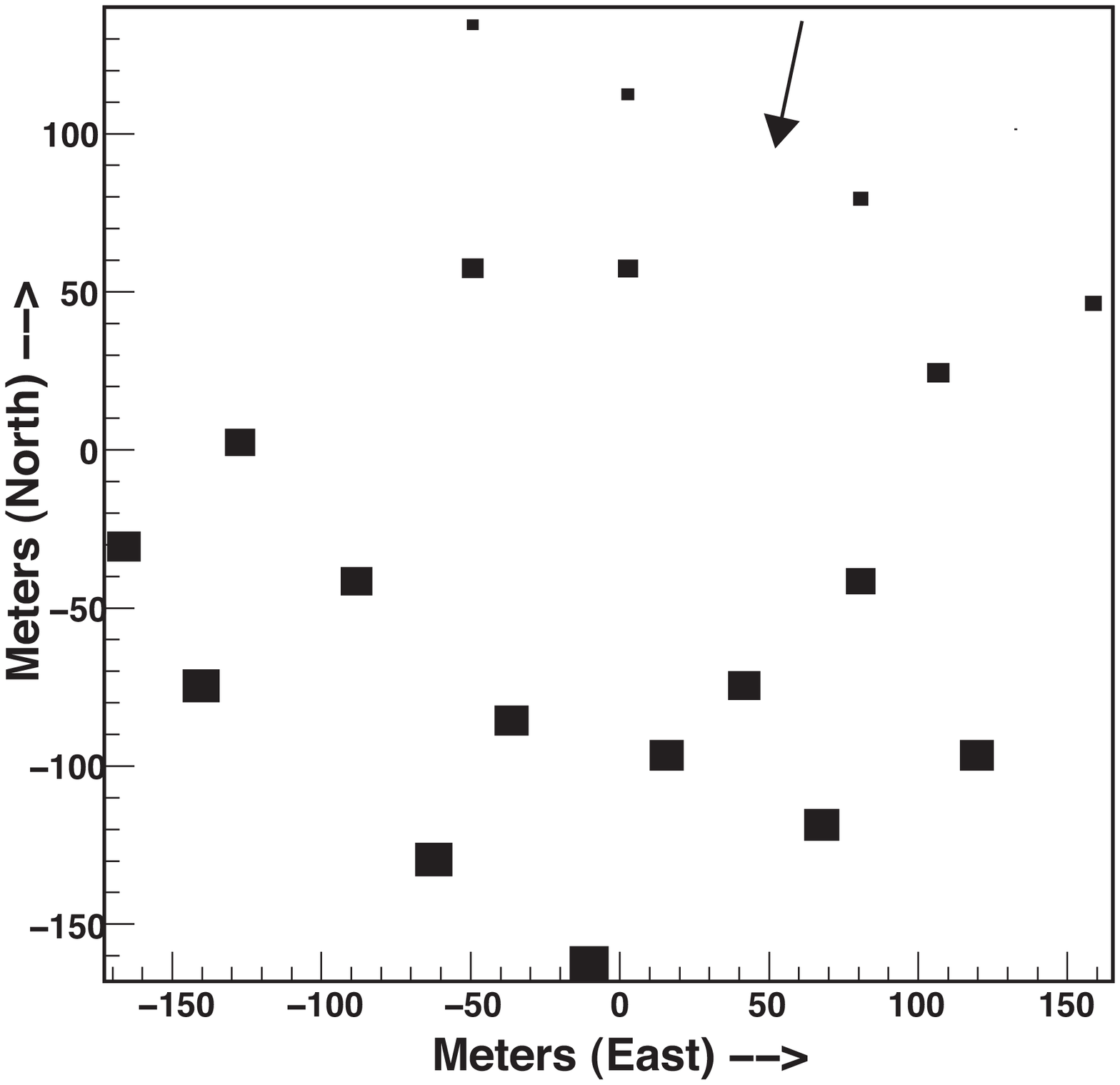}
\hspace{\fill}
\includegraphics[width=\columnwidth]{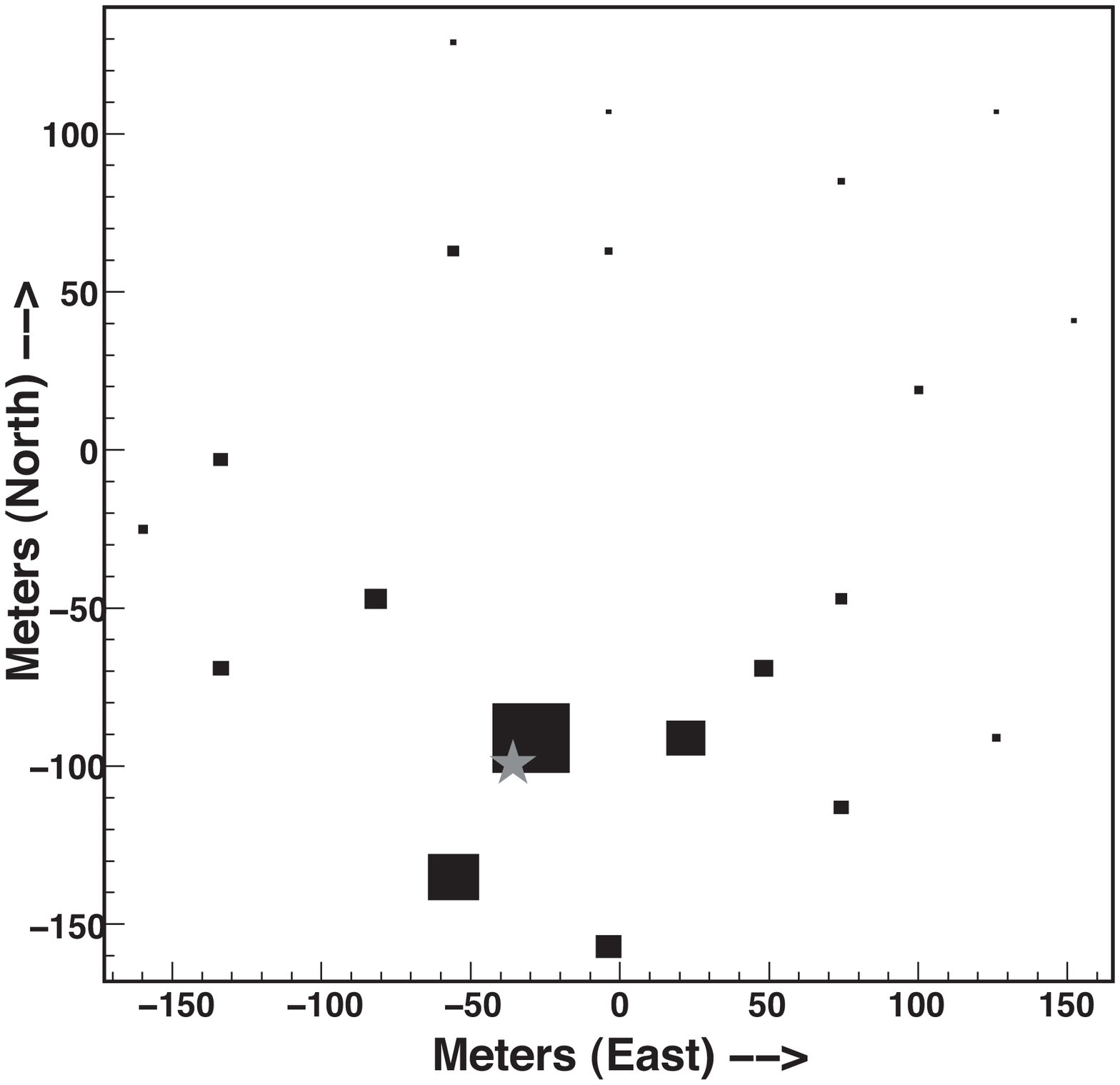}
\caption{\label {event-display} A measured air shower. \LLeft: Arrival
times. \RRight: Energy depositions. The position of each box denotes the
position of a detector on the ground. 
The size of the symbols are proportional to the arrival time (\lleft) and the
energy deposition (\rright).
The reconstructed direction is
$(\theta,\phi)=(17.8\deg\pm0.9\deg, 13.7\deg\pm2.9\deg)$ and the reconstructed
position of the shower axis is $(x_c,y_c)=(-35.7\pm1.6, -99.9\pm1.0)$~m. The
arrow in the \lleft\ panel indicates the azimuth angle $\phi$ of the arrival
direction and the star in the \rright\ panel indicates the position of the
shower axis on the ground.}
\end{figure*} 

\section{Reconstruction of air shower parameters}\label{sect-4}
For a registered air shower, the arrival time of the first particle and the
energy deposition in each detector are measured. Air shower properties are
derived from these quantities, in particular: the arrival time of the shower,
the direction and the position of the shower axis, the lateral density
distribution of charged particles, and the total number of charged particles
contained in the shower. The latter is used to estimate the energy of the
shower-inducing primary particle.

An example of a measured air shower is given in \fref{event-display}. The
left panel represents the measured arrival times and the right panel
shows the energy depositions in the detectors. 

\subsection{Arrival direction}
The arrival time of the particles in the detector is taken as the time at which
the recorded signal crosses the threshold. The measured values are corrected
for time offsets from different electron transit times in the photomultiplier
tubes, different signal propagation speeds in different electronic channels,
and different signal cable lengths.  The average offset for each detector is
determined from the distribution of the differences of the air shower arrival
times in two detectors. The main time offsets result from different cable
lengths.

Using the relative signal arrival times between the detectors, the arrival
direction of the primary cosmic ray is reconstructed.  We assume that the air
shower particles move in a plane towards the ground and we  neglect the small,
but finite curvature of the shower front \cite{Gla2005}. We assume this plane
moves with the speed of light $c$ in the direction of its normal towards the
ground. The normal to the shower plane is taken as the arrival direction of the
primary cosmic ray (or the direction of the shower axis). 

The direction of the shower axis is calculated by minimizing the function
\begin{equation}\label{eq-chi2}
 \delta^2=\sum_{i=1}^k \left[lx_i+my_i+nz_i+c(t_i-t_0)\right]^2,
\end{equation} 
where the summation is over the total number of detectors $k$, $(x_i,y_i,z_i)$
denote the position of the $i^{th}$ detector on the ground, $t_i$ the relative
signal arrival time in that detector, measured with respect to the first hit
detector, and $t_0$ denotes the time at which the shower plane passes through
the origin of the coordinate system. The origin is taken as the center of the
LORA detector array. $(l,m,n)$ denote the direction cosines of the normal
to the plane and are related to the orientation of the shower axis.

Minimizing \eref{eq-chi2}, we obtain the best fit values of $(l,m,n,t_0)$.
The zenith angle of the shower axis, measured from the vertical direction is
obtained as,
\begin{equation}
\theta=\mathrm{sin^{-1}}\left(\sqrt{l^2+m^2}\right) ,
\end{equation}
and the azimuthal angle, measured clockwise from the North through East is obtained using
\begin{equation}
\phi=\mathrm{cos^{-1}}\left(\frac{m}{\sqrt{l^2+m^2}}\right) .
\end{equation} 

\subsection{Position of shower axis and lateral density distribution}
The measured energy deposition in each detector is corrected for the increase
in track length inside the detector due to the inclination angle of the air shower by multiplying by a $\cos(\theta)$ factor. 
Dividing the amount of energy deposition in each detector
by the mean energy deposit per particle, delivers the number of charged particles,
hitting the detector. 

The particle density $n_i$ in each detector is calculated by dividing the
measured number of particles by the effective detector area $A_d \cos
\theta$.  $A_d$ denotes the actual geometrical area of the detector and the
factor $\cos\theta$ takes into account the reduction in the effective
area of the detector for inclined showers with zenith angle $\theta$. 

Plotting the measured particle density in the shower plane as function of the distance 
to the shower axis yields the
lateral density distribution.

\begin{figure}[t]
\includegraphics[width=\columnwidth]{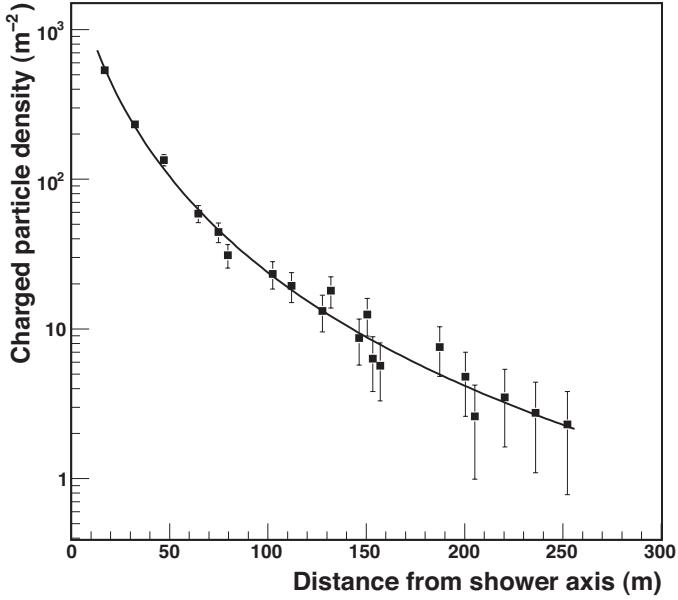}
\caption{\label {event-lateral-density} Lateral density distribution of a 
measured air shower. The solid line represents a fit using the NKG function
\eref{eq-lateral-density} to the reconstructed particle densities in the shower plane. The fit parameters obtained are
$N_{ch}=(5.5\pm0.2)\times 10^6$, $r_M=37.2\pm2.3$~m, with a constant value of
$s=1.7$.}
\end{figure}

The lateral density distribution can be described by a Nishimura-Kamata-Greisen function (NKG) \cite{Ka1958, Gr1960}, given as
\begin{equation}\label{eq-lateral-density}
\rho(r)=N_{ch} C(s)
\left(\frac{r}{r_M}\right)^{s-2}\left(1+\frac{r}{r_M}\right)^{s-4.5},
\end{equation}
where $r$ denotes the radial distance from the shower axis,
$N_{ch}$ is the effective number of charged particles, $r_M$ is the Moli\`ere
radius, $s$ is the lateral shape parameter
(frequently referred to as the ``shower age"). $C(s)$ is given as
\begin{equation}\label{eq-Cs}
C(s)=\frac{\Gamma(4.5-s)}{2\pi r_M^2 \Gamma(s)\Gamma(4.5-2s)} .
\end{equation}
Since the detectors are also sensitive to the photon component of air showers
(see \sref{sect-3}), $N_{ch}$ is an effective number of charged particles,
containing also a fraction of converted photons.
The position of the shower axis on the ground corresponds to the position where
the primary cosmic ray would have hit, if it would not have interacted with the
Earth's atmosphere. By fitting a NKG function to the measured density distribution, the position of the shower axis and the total number of
charged particles can be determined simultaneously along with the other two
parameters $r_M$ and $s$.
The NKG function has been originally derived for electromagnetic cascades with a
constant $r_M$. However, in practice the NKG function is also used to fit
lateral distributions of the hadronic or muonic component of extensive air
showers, see e.g. \cite{An2001}. In such a case, $r_M$ is usually treated as a
free parameter, its value depends on the particle types a detector is sensitive
to (electrons, converted photons, muons, hadrons) and the detection thresholds
for the various particle species.

For the reconstruction of the shower parameters, it is convenient to transform
the detector coordinates into the shower frame of reference. The origin of the
shower frame is taken as the center of the detector array with the z-axis taken
along the shower axis and the x-y plane containing the shower plane. 
The reconstruction needs to be done in several
steps. First, proper starting values of the air shower parameters need to be
provided to initiate the minimization procedure.
For the position of the shower axis $(X_c,Y_c)$ in the shower frame, a good
starting value can be obtained, using the center of gravity of the energy
depositions 
\begin{equation}\label{eq-XcYc}
 X_c=\frac{{\sum\limits^{4}_{i=1}} X_i n_i}{\sum\limits^{4}_{i=1} n_i}\; ;
 \quad Y_c=\frac{\sum\limits^{4}_{i=1} Y_i n_i}{\sum\limits^{4}_{i=1} n_i} ,
\end{equation}
where $(X_i,Y_i)$ denote the coordinates of the detectors in the shower frame
and the summation is over the 4 detectors which recorded the highest energy
depositions.

Using  \eref{eq-lateral-density}, the measured lateral density at the position
of the $i^{th}$ detector $(X_i,Y_i)$ in the shower frame can be written as 
\begin{equation}\label{eq-ni}
  n_i(X_i,Y_i)=N_{ch} F_i(X_i,Y_i) .
\end{equation}
$F_i(X_i,Y_i)$ represents the normalized lateral density distribution function
\begin{equation}
  F_i(X_i,Y_i)=C(s)
  \left(\frac{r_i}{r_M}\right)^{s-2}\left(1+\frac{r_i}{r_M}\right)^{s-4.5} ,
\end{equation}
where $r_i=\sqrt{(X_c-X_i)^2+(Y_c-Y_i)^2}$ is the distance of $(X_i,Y_i)$ from
the position of the shower axis $(X_c,Y_c)$ in the shower frame. Summing
\eref{eq-ni} over the number of detectors, we obtain
\begin{equation}\label{eq-sum-ni}
  \sum\limits^k_{i=1}n_i(X_i,Y_i)=N_{ch} \sum\limits^k_{i=1}F_i(X_i,Y_i) .
\end{equation}
We determine the staring value of $N_{ch}$ using \eref{eq-sum-ni}.  Averaged
values of $r_M$ and $s$ for the measured showers are $r_M=30$~m and $s=1.7$. We
take this value for $r_M$ as initial value for the fit, while $s$ is kept
constant throughout the minimization process.  Fitting $r_M$ and $s$
simultaneously is known to give poor results because of the strong correlation
between them. Simulation studies with CORSIKA have shown that fixing $s$ gives
better results than fixing $r_M$ (see \cite{An2001}). We have checked that
choosing the starting value of $r_M$ in the range of $(20-90)$~m produces
almost the same final values of the fit parameters.

In the first minimization step, we fix $r_M$ and keep the others
$(X_c,Y_c,N_{ch})$ as free parameters. In the second step, we take the results
given by the first fit as starting values. Then, we fix $(X_c,Y_c)$ and
fit the parameters $(r_M,N_{ch})$. These minimization steps are typically iterated three
times with the outputs of each iteration, taken as the starting values for the
next iteration. For the position of the shower axis, the
result of the last iteration is taken as final value. This value can be further
transformed into the position of the shower axis on the ground $(x_c,y_c)$,
following a proper coordinate transformation. For $N_{ch}$ and $r_M$, we go one
step further.  After the last iteration, we determine the lateral density
distribution as a function of the radial distance from the position of the
shower axis. Then, we fit the measured lateral distribution with
\eref{eq-lateral-density} and determine the final values of $N_{ch}$ and $r_M$.

The reconstructed quantities are shown in \fref{event-display} for
illustration. The reconstructed azimuth angle $\phi$ is indicated by the arrow
on the left panel. The star in the right panel represents the reconstructed
position of the shower axis on the ground.  The lateral distribution of the
event shown in \fref{event-display} is depicted in
\fref{event-lateral-density}.  The reconstructed number of charged particles
and the radius parameter are found to be $N_{ch}=(5.5\pm0.2)\times 10^6$ and
$r_M=37.2\pm2.3$~m, respectively.

\begin{figure*}[t]  \centering
\includegraphics[width=\columnwidth]{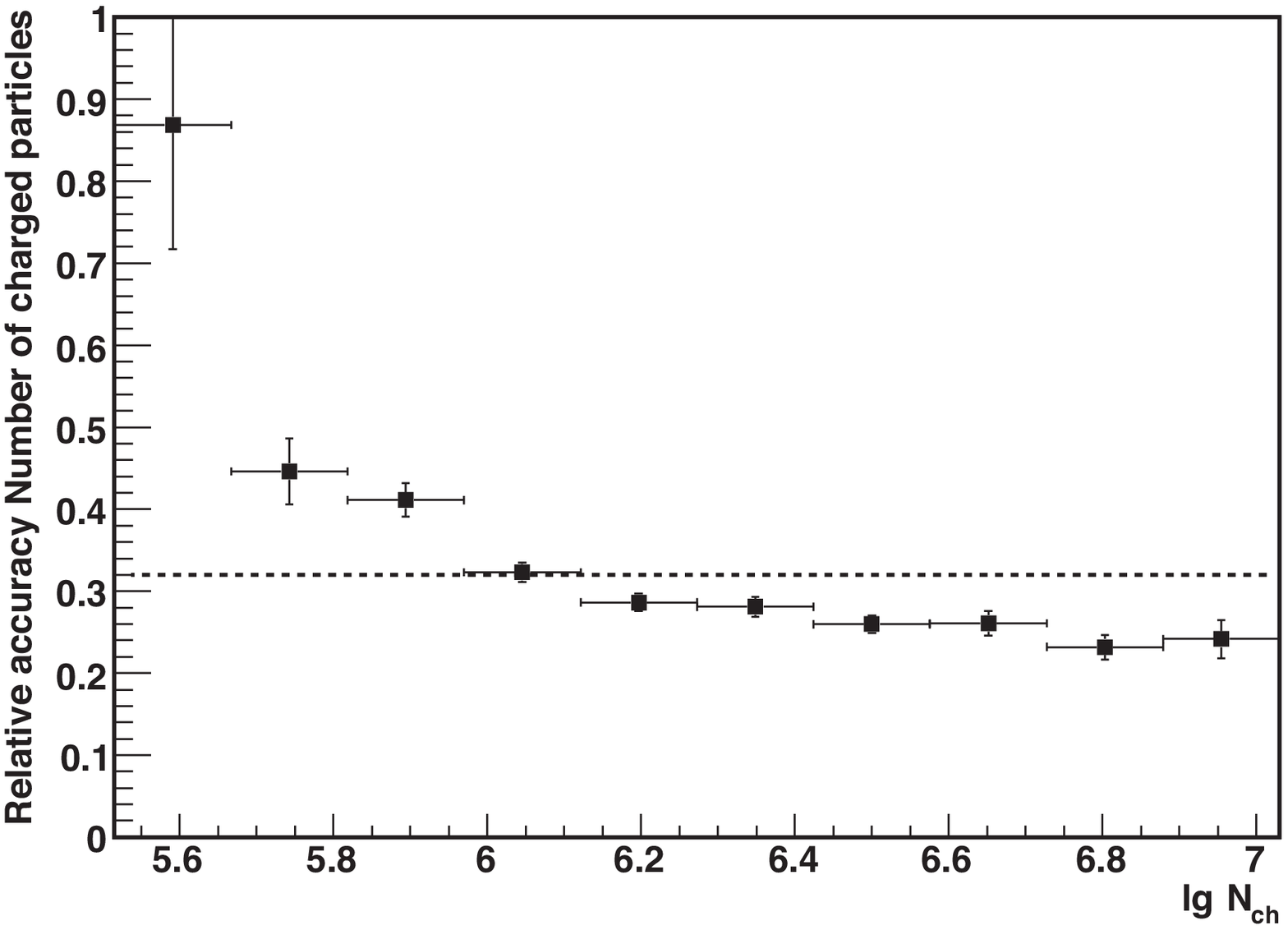}
\hspace{\fill}
\includegraphics[width=\columnwidth]{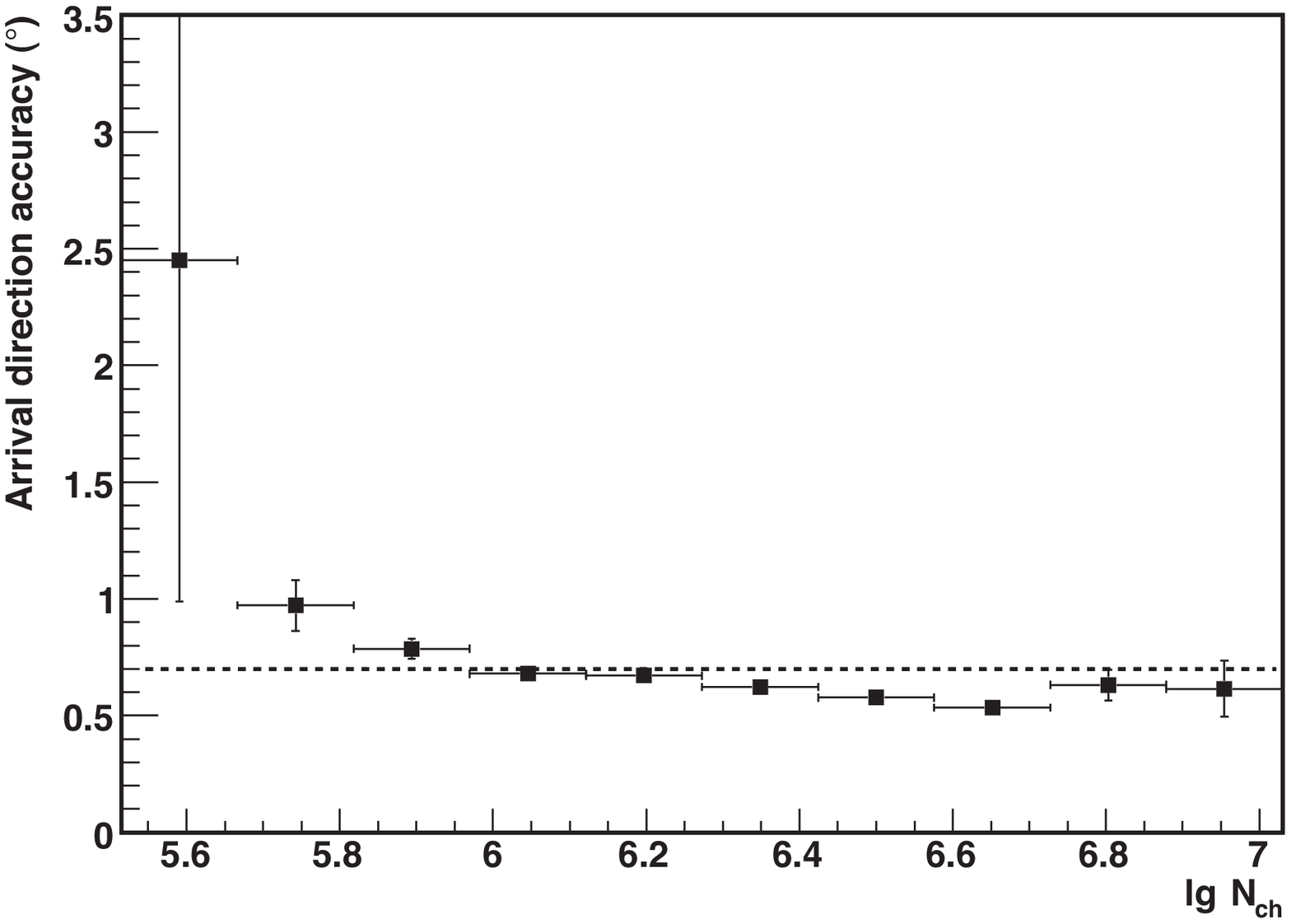}\\
\includegraphics[width=\columnwidth]{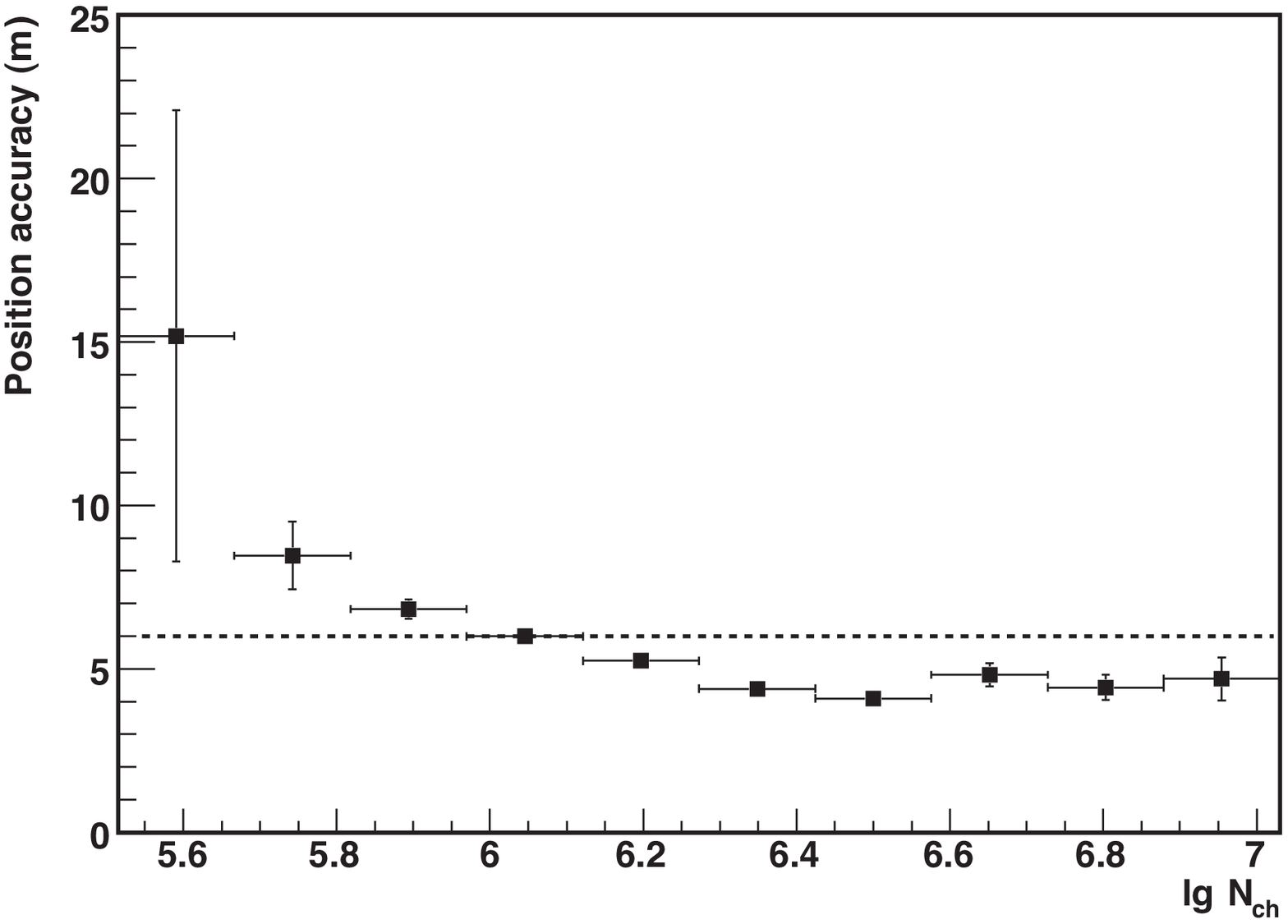}
\caption{\label {reconstruction-accuracies} Measured reconstruction accuracies:
 number of charged particles $N_{ch}$ (\lleft), arrival direction (\rright) and
position of the shower axis on the ground (\bbottom) as a function of the
number of charged particles in a shower. 
For showers with $\lg N_{ch}\gtrsim 6$, the
accuracies are $\lesssim 32\%$ for the number of particles, $\lesssim
0.7^\circ$ for the arrival direction, and $\lesssim 6$ m for the position of
the shower axis. These values are indicated by the dashed lines in the figures.}
\end{figure*}

\begin{table}
\caption{Selection criteria applied during data taking and analysis.}
\vspace{\baselineskip}
\label{cuttab}
\begin{tabular}{ll} \hline\hline
 selection criteria: & \\
 trigger: & 3/4 detectors in an electronics unit \\
 analysis: & 5 detectors with $\ge1$ particle each\\
 quality criteria: & \\
 zenith angle: & $\theta<35^\circ$ \\
 shower axis: & $<150$~m from center of LORA \\
 radius parameter: & $10~\mbox{m} < r_M < 200~\mbox{m}$ \\
 \hline\hline
\end{tabular} 
\end{table}

\begin{figure*}[t]
\includegraphics[width=\columnwidth]{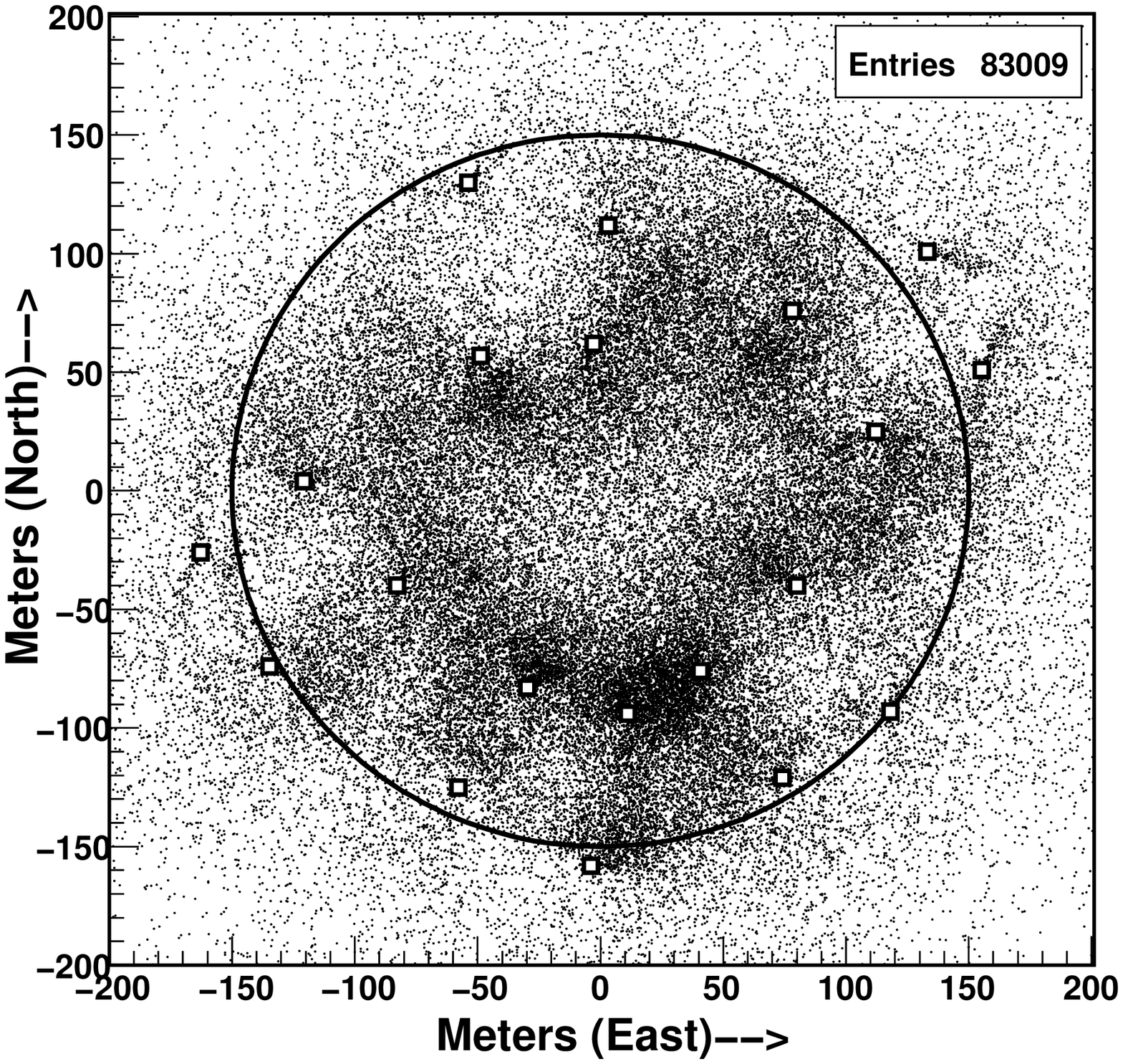}\hspace*{\fill}
\includegraphics[width=\columnwidth]{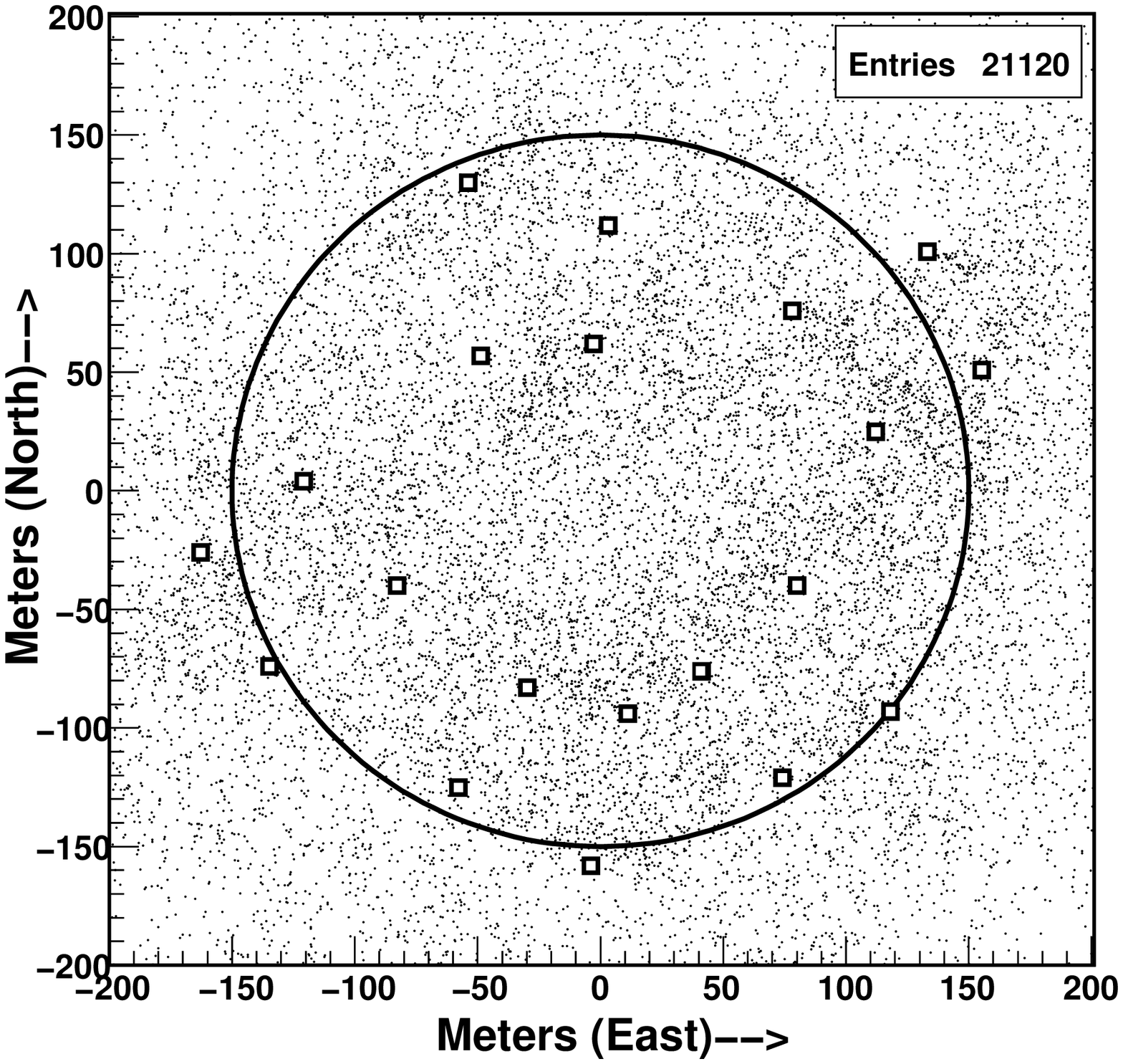}
\caption{\label {core-distribution} Distribution of the reconstructed positions
of the axis for measured air showers. 
\LLeft: all showers. \RRight: only showers with a number of charged particles
$\lg N_{ch}>6.25$. Only showers with characteristic radii in the range
$10~\mbox{m}~\le r_M\le 200$~m are included in the distribution. The open
squares denote detector positions and the circle represents a fiducial area of
radius 150~m around the center of LORA, chosen for our analysis.}
\end{figure*}

\section{Reconstruction accuracies}\label{sect-5}
The accuracies in the reconstruction of air shower parameters are determined
from the data itself using the divided-array method. We divide the full array
into two arrays with larger spacing and compare the reconstructed parameters of the two parts. 
This technique has its limitations as it will not be sensitive to systematic biases and an additional penalty is applied
by reducing the number of measurements per event. However, it gives a good impression of the capabilities of the array. 
A study based on air shower simulations is currently underway. 
This will give a more accurate picture, which is needed for more in-depth analyses of the data.

For the accuracy in the position of the shower axis, we
calculate the difference between the estimates of the two half arrays 
\begin{equation}
\Delta_{pos}^{12}=\sqrt{(x_1-x_2)^2+(y_1-y_2)^2} ,
\end{equation}
where $(x_1,y_1)$ and $(x_2,y_2)$ are the reconstructed positions of the shower
axis on the ground with the two sub-arrays, respectively. Then, the
reconstruction accuracy for the full array $\sigma_{pos}$ is calculated as,
\begin{equation}\label{eq-sigma-core}
\sigma_{pos}=\frac{\sigma^{12}_{pos}}{\sqrt{2}}
\end{equation}
where $\sigma^{12}_{pos}$ denote the spread of the distribution of
$\Delta^{12}_{pos}$.

For the arrival direction accuracy, we calculate the space angle difference between the estimates of the two sub-arrays as,
\begin{equation}
\Delta^{12}_{angle}=\mathrm{cos}^{-1}\left[\mathrm{sin}\theta_1\mathrm{sin}\theta_2\mathrm{cos}(\phi_1-\phi_2)+\mathrm{cos}\theta_1\mathrm{cos}\theta_2\right]
\end{equation}
where $(\theta_1,\phi_1)$ and $(\theta_2,\phi_2)$ are the arrival directions
(zenith and azimuth angle) reconstructed with the two sub-arrays. For the
accuracy in the number of charged particles, the difference between the two
sub-arrays is calculated relative to the number given by the full array as,
\begin{equation}
\Delta^{12}_{ch}=\left(\frac{N^1_{ch}-N^2_{ch}}{N_{ch}}\right)
\end{equation}
where $N^1_{ch}$, $N^2_{ch}$ are the number of particles given by the two
sub-arrays and $N_{ch}$ is the number given by the full array. The
reconstruction accuracies for the arrival direction and the number of particles
are then calculated using a similar relation as given by 
\eref{eq-sigma-core}.

\Fref{reconstruction-accuracies} shows the reconstruction accuracies, derived
from the data as a function of the number of particles determined from the full
array, see \tref{cuttab} for the applied quality criteria.  The panels in the
figure correspond to the number of particles, arrival direction, and the
position of the shower axis, respectively. The reconstruction accuracies for
showers with a number of charged particles $\lg N_{ch}\gtrsim 6$ are
within approximately $32\%$ for the number of charged particles, $0.7^\circ$
for the arrival direction, and 6~m for the position of the shower axis on the
ground.

\begin{figure*}[t]
\includegraphics[width=\columnwidth]{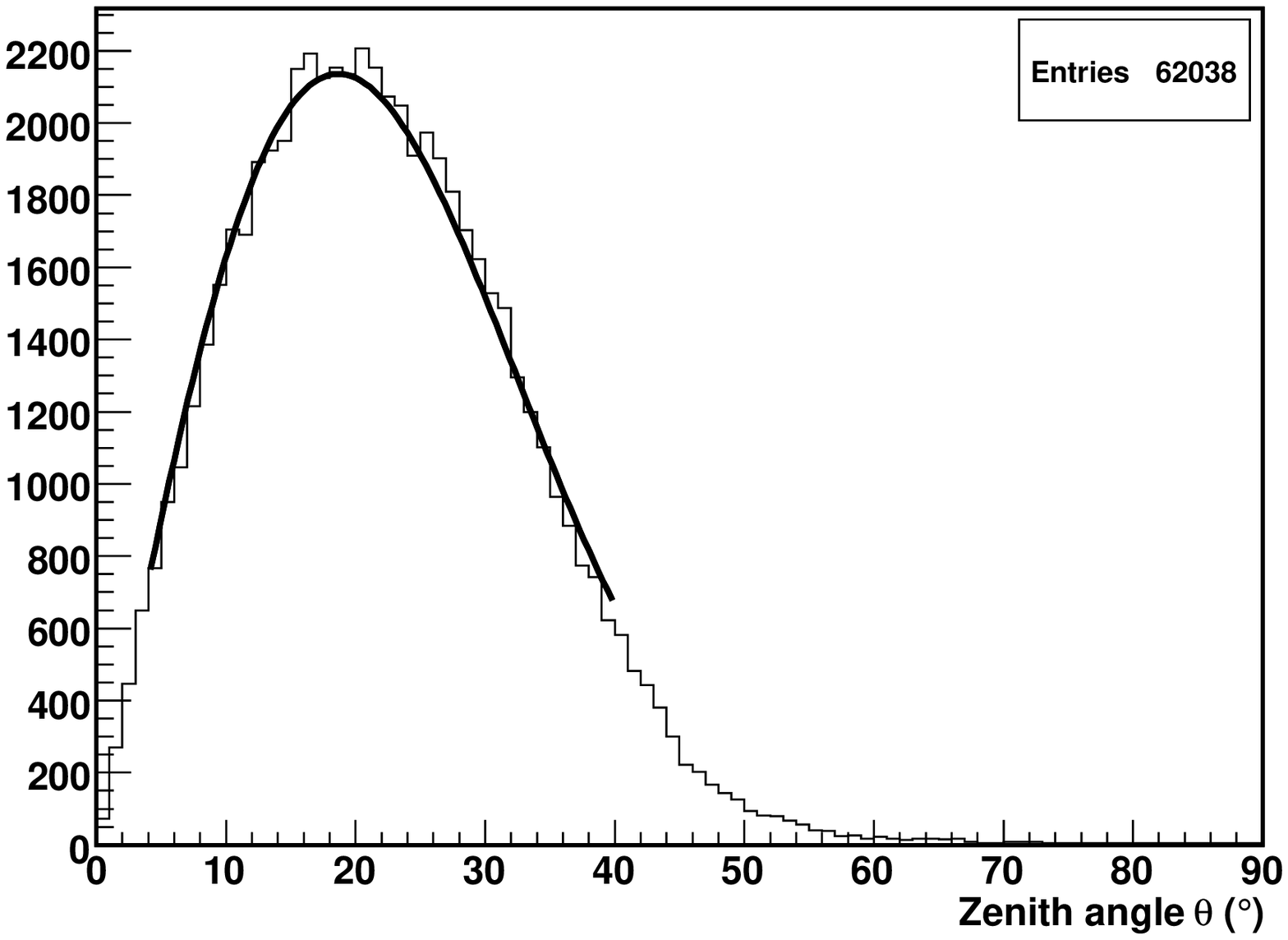}
\hspace{\fill}
\includegraphics[width=\columnwidth]{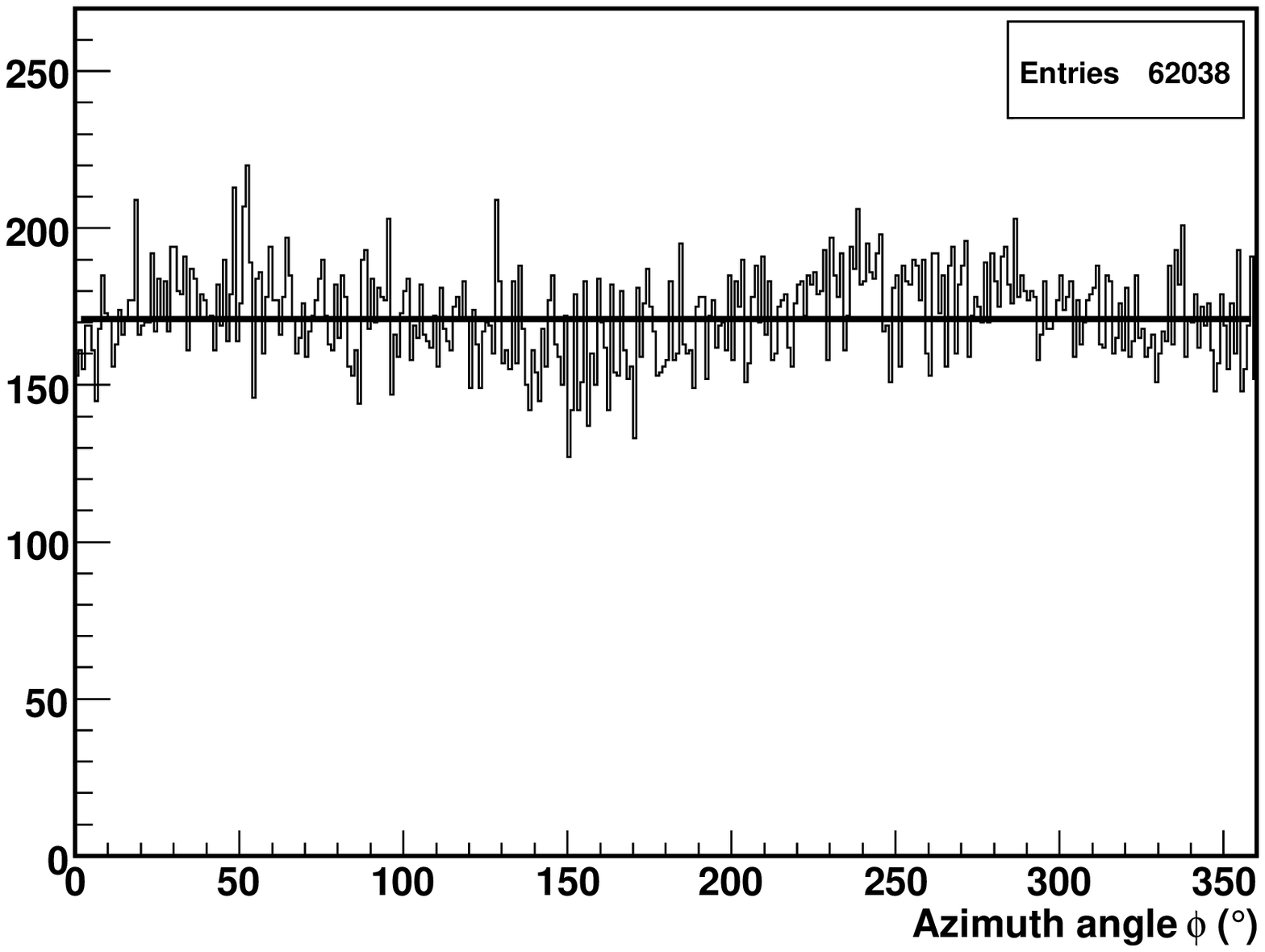}
\caption{\label {zenith-azimuth} Distribution of angles of the arrival direction of
air showers measured with positions of the shower axis within the fiducial area 
(within 150~m from the center of LOFAR) and
characteristic radii in the range $10~\mbox{m}~\le r_M\le 200$~m. \LLeft: Zenith
angle distribution. The thick line is a fit using \eref{zenitheq} in the range
of $4^\circ\leq\theta\leq40^\circ$. The fit parameters are found to be
$a_1=(1.073\times 10^4 \pm 71)$ and $a_2=(8.786\pm 0.062)$. \RRight: Azimuth
distribution. The azimuth is measured eastwards from the north. The horizontal
line represents a straight line fit to the distribution.}
\end{figure*} 

\section{Performance of the array}\label{sect-6}
For the following analysis air showers have been selected that trigger at least
5 detectors with a minimum particle density of 1 particle m$^{-2}$ and which
have characteristic radii in the range of $(10-200)$~m, see also \tref{cuttab}.
The spatial distribution of the reconstructed shower axes is depicted in
\fref{core-distribution}.  The left-hand panel shows the positions for all
showers. On the right-hand side, only showers with a total number of charged
particles $\lg N_{ch}>6.25$ are selected. The full set is dominated by low-energy showers, which show
inhomogeneities, caused by the non-regular spatial arrangement of
the scintillation detectors (left-hand side).  This threshold effect vanishes at higher energies, which correspond to larger values of $N_{ch}$. Above $\lg N_{ch}=6.25$ one recognizes that the
showers are reconstructed more uniformly. The squares in the figure
represent detector positions and the circle represents a fiducial area of
radius 150~m around the array center.  For our analysis in the following, we
only select those air showers with position of the shower axis falling within
this fiducial area. This area is chosen such that we include in our analysis
only those air showers with reliable estimates of the position of the shower
axis and at the same time, retain as many showers as possible. 

\Fref{zenith-azimuth} shows the zenith angle  and the azimuthal angle
distribution of the measured air showers.  The zenith angle distribution
is fitted in the range of $(4^\circ-40^\circ)$ with the distribution function 
\begin{equation} \label{zenitheq}
  f(\theta) \,\mbox{d}\theta =a_1\; \sin\theta \cos^{a_2}\theta 
    \,\mbox{d}\theta .
\end{equation}
The fit parameters are found to be $a_1=(1.073\times 10^4 \pm 71)$ and
$a_2=(8.786\pm 0.062)$. The thick curve in the  figure represents the fit
result. The peak of the distribution is found to be at $\theta\sim 19^\circ$.
The steep rise in the distribution below the peak is due to the increase in the
solid angle with the zenith angle. The steep fall above the peak is due the
combined effect of the decrease in the effective collection area of the array
and the increase in the shower attenuation at larger zenith angles. The effect
of the attenuation is expected to be more significant for showers initiated by
low-energy primaries.

The azimuth distribution is almost uniform at all angles. This is expected
because of the high level of isotropy in the arrival directions of cosmic rays of this energy range
which is related to their diffusive nature of propagation in the Galaxy. 
However, there is some structure visible. This is due to the irregular positioning of the detectors. 
For different azimuth angles the projected distances between detectors get smaller or wider, 
which affects the reconstruction. The structure is, however, not severely affecting the uniformity. 
This is illustrated by the horizontal line in \fref{zenith-azimuth}, which represents a
straight line fit to the measured distribution, illustrating what equally distributed azimuth angles would look like. 
This fit shows a $\chi^2/\mathrm{ndof}$ of close to one.

\section{Conclusions}
LORA is an air shower array that has been built for cosmic-ray measurements
with LOFAR. Its primary purpose is to trigger the read-out of the LOFAR radio
antennas for cosmic-ray events and to provide basic air shower properties, such
as the position of the shower axis, the arrival direction, and the energy of
the primary particle. The full set-up of the LORA array was completed in June
2011. It currently operates as standard triggering tool for the air shower detection with LOFAR. 

The array is comprised of 20 scintillation detectors and measures the arrival
direction of high-energy air showers (with $\lg N_{ch}>6$) on average with an
accuracy better than $0.7^\circ$, the position of the shower axis better than
6~m, and the number of charged particles better than 32\%.

The air-shower information determined by LORA is used as input for the
reconstruction of air shower properties with the LOFAR radio antennas. The
measured air showers are also used to optimize a radio-only trigger for LOFAR.

\section*{Acknowledgment}
We are grateful for technical support from ASTRON. In particular, we would like
to thank
J.~Nijboer,
M.~Norden,
K.~Stuurwold
and H.~Meulman 
for their support during the installation of LORA in the LOFAR core. We are grateful to the KASCADE-Grande collaboration, who generously lent
the scintillator units to us.

We acknowledge funding from the Samenwerkingsverband Noord-Nederland (SNN), the Netherlands Research School for Astronomy (NOVA) and
from the European Research Council (ERC).
LOFAR, the Low Frequency Array designed and constructed by ASTRON, has facilities in several countries, that are owned by various parties (each with their own funding sources), and that are collectively operated by the International LOFAR Tele- scope (ILT) foundation under a joint scientific policy.

\end{document}